\documentclass[pre,aps,showpacs,amsmath,amssymb]{revtex4-1}
\usepackage{graphicx}
\usepackage{bm}
\begin{document}

\title{Interfacial instabilities in two-dimensional Stokes flow: a weakly nonlinear analysis}
\author{Gabriel D. Carvalho$^{1}$}
\email[]{gabrieldiascarvalho@outlook.com}
\author{Rodolfo Brand\~ao$^{2}$}
\email[]{r.brandao18@imperial.ac.uk}
\author{Jos\'e A. Miranda$^{1}$}
\email[]{jme@df.ufpe.br}
\affiliation{$^1$ Departamento de F\'{\i}sica, Universidade Federal de Pernambuco, Recife, Pernambuco  50670-901 Brazil \\ 
$^2$ Department of Mathematics, Imperial College London, London SW7 2AZ United Kingdom}

\begin{abstract}
Two-dimensional Stokes flow with injection and suction is investigated through a second-order, perturbative mode-coupling approach. We examine the time-dependent 
disturbance of an initially circular interface separating two viscous fluids, and derive a system of nonlinear differential equations describing the evolution of the 
interfacial perturbation amplitudes. Linear stability analysis reveals that an injection-induced expanding interface is stable, while a contracting motion driven by 
suction is unstable. Curiously, at the linear level this suction instability is independent of the viscosity contrast between the fluids. However, second-order results tell a different story, and show that the viscosity contrast plays a key role in determining the shape of the emerging interfacial patterns. By focusing on the onset of nonlinear effects, we have been able to extract valuable information about the most fundamental features of the pattern-forming structures for arbitrary values of viscosity contrast and surface tension.\end{abstract}

\maketitle

\section{Introduction}
\label{intro}

Complex pattern formation flourishes in nature and has been vigorously investigated in a number of physical, chemical, and biological systems~\cite{Pelce,Mein,Cross}. 
In this broad field of scientific research, one chief point of interest is to try to understand the morphology of the patterns that emerge at the interface separating two different 
phases. In particular, the development of interfacial instabilities and patterns is an appealing problem in fluid dynamics. The study of pattern-forming structures arising 
in systems such as Taylor-Couette flow~\cite{Ahl}, Rayleigh-B\'enard convection~\cite{Man}, and Rayleigh-Taylor instability~\cite{Limat} has motivated numerous experimental
and theoretical works. 

An important example of interfacial pattern formation in fluid systems is the celebrated Saffman-Taylor (or, viscous fingering) problem~\cite{Saf}. It considers the 
occurrence of interfacial instabilities when a fluid displaces another of higher viscosity between the narrowly spaced plates of a Hele-Shaw cell~\cite{Rev}. Under such circumstances, 
the fluid-fluid interface is unstable leading to the growth of fingerlike shapes. The formation of these viscosity-difference-driven fingering shapes is the result of the interplay 
between destabilizing pressure gradients, and stabilizing surface tension effects. In this context, the dynamic evolution of the viscous fingers is described by an effectively 
two-dimensional, gap-averaged Darcy's law, which connects the fluid velocity to the pressure gradient. 

For radial fluid injection~\cite{Pat,Chen,Cardoso,Mir1,Praud,Li,Bischo}, the less viscous fluid is injected at the center of the Hele-Shaw cell, and drives radially the more viscous 
fluid. This makes the expanding fluid-fluid interface to deform, leading to the rising of interfacial fingers. The growing structures tend to bifurcate at their tips, and multiply through a repeated 
tip-splitting process. The result is the formation of complex, highly branched patterns. On the other hand, for the equivalent problem with radial fluid suction~\cite{Pat,Thome,chensuc}, 
where a blob of a more viscous fluid (surrounded by a less viscous fluid) is sucked radially inward into a sink located at the center of the cell, the fingering formation is a bit 
different. During the suction-driven flow, the initially circular fluid-fluid interface shrinks and deforms by the penetration of multiple fingers of the less viscous outer fluid. 
Ultimately, a single finger reaches the sink, while the remaining fingers essentially stop their inward moving growth. Nevertheless, under suction, the penetrating fingers are quite smooth, 
and do not tend to split at their tips. The zero and small surface tension limits of both injection and suction cases have also been studied via conformal mapping methods and numerical simulations, where researchers have identified the development of a finite-time blow up of the interfacial solutions, as well as cusp formation phenomena~\cite{Rich1,How1,How2,Rich2,Cum,Ceni1,Ceni2}. 

It should be noted that, in the framework of the Saffman-Taylor problem (under either injection or suction) there is no instability when a more viscous fluid displaces 
a less viscous one, constituting a stable, reverse flow displacement. Likewise, a Hele-Shaw flow with viscosity-matched fluids is also stable. In these cases, the fluid-fluid 
interface propagates axisymmetrically as a stable circular front. Therefore, within the scope of the Darcy's law-regulated Saffman-Taylor problem, the interface instability 
is highly dependent on the viscosity contrast (or the dimensionless viscosity difference)
\begin{equation}
\label{contrast}
A=\frac{\eta_{2} - \eta_{1}}{\eta_{2} + \eta_{1}},
\end{equation}
where $\eta_{2}$ ($\eta_{1}$) denotes the viscosity of the outer (inner) fluid, and $-1 \le A \le 1$. For the Saffman-Taylor problem under injection (suction), the interface is unstable (stable) 
for $A >0$ ($A \ge 0$), and stable (unstable) for $-1 \le A \le 0$ ($-1 \le A < 0$).

Curiously, the equivalent injection and suction pattern-forming problems for two-dimensional Stokes flow have received less attention~\cite{Tanv1,Tanv2,Poz,How3,Cum2,Amar1,Amar2}. Two 
possible reasons for this fact are (i) the practical difficulty to implement a legitimate two-dimensional Stokes flow experimentally, and (ii) the greater difficulty of the 
Stokes flow calculations compared to the Darcy's flow ones. However, the studies performed in Refs.~\cite{Tanv1,Tanv2,Poz,How3,Cum2,Amar1,Amar2} 
reveal that  the two-dimensional Stokes flow instability for injection and suction differ significantly from their Darcy's flow, Saffman-Taylor counterparts. 

Analytical and numerical solutions of the time-evolving two-dimensional Stokes flow problem based on powerful complex variable methods~\cite{Tanv1,Tanv2,Poz,How3,Cum2,Amar1,Amar2} show that 
an injection-driven expanding bubble (bubble of negligible viscosity displacing a viscous outer fluid) is stable.  It has been found that a growing bubble will approach an 
expanding circle for later times. On the other hand, it has been shown that a suction-driven, contracting circular bubble is unstable to disturbances. These observations 
are true both in the presence, or (even) in the absence of surface tension.  In the absence of surface tension, the solutions for the interface under suction will in general break-down in a 
finite time, owing to the formation of cusp singularities on the bubble surface. In contrast, suction of a bubble under finite surface tension sets a time scale for which 
narrow structures (known as ``almost-cups", or ``near-cusps") can develop. The greater the surface tension, the later the near-cusp will appear. Analogous results have been observed 
in Refs.~\cite{How3,Cum2} for the suction of a blob of viscous fluid surrounded by an outer fluid of negligible viscosity under two-dimensional Stokes flow circumstances. 

Nonetheless, the most striking and somewhat surprising results achieved in Refs.~\cite{Tanv1,Tanv2,Poz,How3,Cum2,Amar1,Amar2} for two-dimensional Stokes flow 
are not those related to the surface tension, but the ones pertaining the effects of the viscosity contrast. Although the complex variable calculations carried out 
in Refs.~\cite{Tanv1,Tanv2,Poz,How3,Cum2} are limited to extreme values of the viscosity contrast (i.e., $A=1$ for the bubble problem, and $A=-1$ for the blob situation), 
their results demonstrate that, as opposed to the Saffman-Taylor case, the two-dimensional Stokes instability for suction occurs independently of the sign of the 
viscosity contrast $A$ between the fluids. Similar findings have also been obtained in the linear stability analyses of the two-dimensional Stokes flow for suction 
performed in Refs.~\cite{Cum2,Amar1,Amar2}. As a matter of fact, the analytical linear stability theory developed in Refs.~\cite{Amar1,Amar2} considers a whole range 
of values for the viscosity contrast (i.e., $-1 \le A \le 1$ ), but found that the linear dispersion relation (linear growth rate) is independent of the sign and magnitude of $A$. This is in stark contrast to what happens in the usual viscous fingering problem with suction~\cite{Pat} where the viscosity contrast plays a very important role already in the linear regime. All these observations highlight the fact that the two-dimensional Stokes flow instability for suction and injection 
differs significantly from the corresponding Saffman-Taylor instability. 

Previously, we have commented that one of the detrimental facts for the relatively lower interest in the study of two-dimensional Stokes flow was the difficulty in implementing 
real two-dimensional experimental realizations of such a fluid system. Nevertheless, the rapid development of microfluidics~\cite{stone}, superhydrophobic surfaces~\cite{Roth}, 
and interesting dynamical experiments in inhomogeneous lipid membranes~\cite{Amar2} recently brought renewed interest in the theory and experiments of two-dimensional Stokes 
flows~\cite{Crow1,Crow2}. One particularly interesting set of works has been recently published in Refs.~\cite{Sayag1,Sayag2,Sayag3}, wherein the authors designed an apparatus allowing a thin and uniform layer of viscous fluid to propagate between two tractionless surfaces. Owing to the absence of wall friction, the flow is vertically uniform and satisfies a radial planar Stokes flow. By conducting advanced time experiments~\cite{Sayag1} and linear stability theory~\cite{Sayag2} in this geometry, the authors showed that the interface between a (non-Newtonian) shear-thinning fluid displacing a lower-viscosity fluid can become unstable. Furthermore, the  experiments also reveal the formation of peculiar fingers exhibiting rectangular-shaped tips. The planar  problem studied in Refs.~\cite{Sayag1,Sayag2} has been extended to a curved geometry environment in Ref.~\cite{Sayag3} where a similar two-dimensional Stokes flow has been examined  on the surface of a sphere. As argued in Ref.~\cite{Sayag3}, the investigation of such instabilities in nonplanar geometries could offer useful insights into practical situations involving the formation of glacier ice rifts, and ice shelves on a planetary scale (icy moons).

The instability found in Refs.~\cite{Sayag1,Sayag2,Sayag3} is fundamentally different from the one that arises in the classical Saffman–Taylor 
viscous fingering problem and its emergence strongly relies on the absence of boundary friction. Therefore, as stated by the authors in Refs.~\cite{Sayag1,Sayag2,Sayag3}, the two-dimensional Stokes flow configuration they  study is similar to flow in a Hele-Shaw apparatus, but with no-stress instead of no-slip boundary conditions along the plates of the cell. As suggested in Ref.~\cite{Sayag3}, one way to reduce traction with the plates in the Hele-Shaw setup could be using superhydrophobic surfaces. This opens up the possibility to perform realistic two-dimensional Stokes flow experiments using the well known and vastly utilized Hele-Shaw cells, which certainly could help with the resurgence of practical interest in two-dimensional Stokes flow.

Motivated by the new possibilities regarding experimental realizations of the two-dimensional Stokes flow, and by the prospects of using it as a theoretical tool 
to examine a diverse spectrum of systems, in this work, we develop the weakly 
nonlinear analysis of the two-dimensional Stokes flow with injection and suction. Existing analytical and numerical treatments of the problem describe 
the early and late time stages of the fluid-fluid interface, in the zero and small surface tension limits, mostly using conformal mapping techniques, or by employing boundary integral 
formulations. In addition, the majority of these studies focus on the cases in which the viscosity contrast is either $A=-1$ or $A=1$, and just a few works address the effect of $A$ 
(with $-1 \le A \le 1$) but are limited to the purely linear regime. Conversely, our work studies the intermediate stage between the linear and the fully nonlinear ones, focusing 
especially on the onset of nonlinear effects. Our approach applies to any value of surface tension and $A$, and gives insight into the mechanisms of pattern formation. The linear stability 
analysis applies only to very early stages of the flow, and does not offer ways to predict the morphological aspects of the essential cusp and near-cusp phenomena detected in later stages of the two-dimensional Stokes flow. Here we show that a perturbative,  second-order mode-coupling theory is able to tackle some of these issues and to investigate the role of $A$ in the cusp-like shapes, an analysis which was not performed in previous studies.

\begin{figure}
\includegraphics[width=3.5 in]{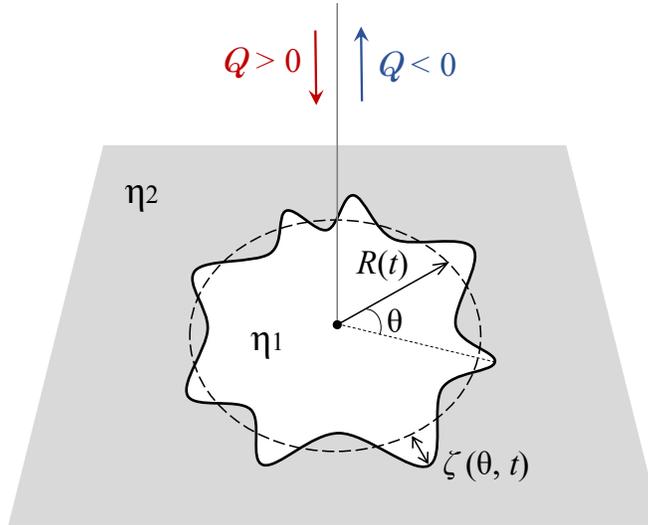}
\caption{A schematic of the two-dimensional Stokes flow problem with injection ($Q>0$) or suction $Q<0$, where $Q$ is the areal rate. The outer (inner) fluid 
	has viscosity $\eta_2$ ($\eta_1$). The time-dependent unperturbed fluid-fluid interface radius is represented by $R(t)$, and the interface 
perturbation is denoted by $\zeta(\theta,t)$, where $\theta$ is the azimuthal angle.}
\label{fig1}
\end{figure}

\section{Basic equations and the weakly nonlinear approach}
\label{derive}

Consider the two-dimensional flow of a two-fluid system where an inner fluid 1 is surrounded by an outer fluid 2. These fluids are immiscible, incompressible, and viscous. Initially, the 
interface separating the fluids is circular and centered at the origin of the coordinate system. Fluid 1 can be injected or sucked at the origin with a constant areal rate $Q$ 
(area covered per unit time). The cases of a single point source ($Q > 0$) and sink ($Q < 0$) correspond to injecting or sucking, respectively. In this framing, fluid flow is governed by the incompressible Stokes equations~\cite{Batchelor,Langlois}
\begin{equation}\label{stokes1}
    \boldsymbol{\nabla} p_{j} = \eta_{j}\nabla^2\mathbf{u}_j,
\end{equation}
and
\begin{equation}\label{stokes2}
  \boldsymbol{\nabla}\cdot \mathbf{u}_j = 0,
\end{equation}
wherein $\eta_j$, $p_j$ and $\mathbf{u}_j$ denote the viscosity, pressure, and velocity fields of fluid $j$, with $j=1 (2)$ labels the inner (outer) fluid.  Equation~(\ref{stokes1}) 
is derived from the Navier-Stokes equation if one omits inertial terms (low Reynolds number limit), while Eq.~(\ref{stokes2}) expresses the incompressibility of the fluids. Additionally, 
we assume that the interface between the fluids has a constant surface tension $\sigma$.

We parameterize the fluid-fluid interface by $r =   \mathcal{R}(\theta, t)$, where $(r,\theta)$ are the usual polar coordinates centered at the injection (or suction) point, and 
define the unit normal and tangent vectors to the interface as
\begin{equation}\label{rt}
    \boldsymbol{\hat{n}} = \frac{\mathcal{R}\mathbf{\hat{r}}  - \mathcal{R}_{\theta}\boldsymbol{\hat{\theta}} }{\left(\mathcal{R}^2 + \mathcal{R}_{\theta}^2\right)^{1/2}}, \qquad     \boldsymbol{\hat{t}} = \frac{\mathcal{R}_{\theta}\mathbf{\hat{r}}  + \mathcal{R}\boldsymbol{\hat{\theta}} }{\left(\mathcal{R}^2 + \mathcal{R}_{\theta}^2\right)^{1/2}},
\end{equation}
where the subscripts denote a derivative with respect to $\theta$. In Eq.~(\ref{rt}), $\mathbf{\hat{r}}$ ($\boldsymbol{\hat{\theta}}$) is the unit vector along the radial (azimuthal) direction.
At the interface, we adopt the usual boundary conditions for two-dimensional Stokes flow~\cite{Batchelor,Langlois,Tanv1,Amar1}, for which velocity is continuous
\begin{equation}\label{bc12}
    \mathbf{u}_1 = \mathbf{u}_2.
\end{equation}
The dynamic condition on the interface is obtained by considering the stress balance between the two fluids in both the normal and tangential directions. The balance
equation of tangential stress is given by
\begin{equation}\label{bc3}
    \boldsymbol{\hat{n}}\cdot \left(\boldsymbol{\pi}_1 - \boldsymbol{\pi}_2\right)|_{ \mathcal{R}}\cdot \boldsymbol{\hat{t}} = 0, 
\end{equation}
establishing the continuity of the shear stress. In addition, the normal stress balance equation gives
\begin{equation}\label{bc4}
    \boldsymbol{\hat{n}}\cdot \left(\boldsymbol{\pi}_1 - \boldsymbol{\pi}_2\right)|_{ \mathcal{R}}\cdot \boldsymbol{\hat{n}} = - \sigma \kappa,
\end{equation}
ensuring that the jump in the normal stress across the interface equals the product of the surface tension $\sigma$ and the curvature of the interface $\kappa$. Since both 
fluids are assumed to be Newtonian, the stresses are related to pressure and velocity as follows~\cite{Batchelor,Langlois}
\begin{equation}
    \boldsymbol{\pi}_j = - p_{j}\mathbf{I} + \eta_{j}\left[ \mathbf{\nabla} \mathbf{u}_j + \left(\mathbf{\nabla} \mathbf{u}_j\right)^{T} \right],
\end{equation}
where $\mathbf{I}$ is the identity matrix, and  the superscript $T$ denotes a matrix transpose. Moreover, since the interface is free, we impose a kinematic velocity condition relating the normal 
velocity on the interface to the interfacial deformation \cite{Tanv1,Mir1}
\begin{equation}\label{kin}
    \frac{\partial \mathcal{R}}{\partial t} = \mathbf{u}_j\cdot \boldsymbol{\hat{n}} |_\mathcal{R}\left(1 + \frac{\mathcal{R}^2_{\theta}}{\mathcal{R}^2}\right)^{1/2}.
\end{equation}
Equation~(\ref{kin}) expresses the fact that the normal velocity of a point on the interface is equal to the normal component of fluid velocity at that point. It is based on the assumption that the fluid-fluid boundary moves along with the fluid particles, coupling the motion of the interface to the motion of the bulk fluids.

After establishing the governing hydrodynamic equations~[Eqs.~(\ref{stokes1})-(\ref{stokes2})] and the relevant boundary conditions~[Eqs.~(\ref{bc12})-(\ref{kin})] of the two-dimensional Stokes 
flow problem, we are ready to begin describing our perturbative, weakly nonlinear approach. However, before moving on, we wish to make a few important remarks. First, as already 
briefly discussed in Sec.~\ref{intro}, it should be stressed that there are important differences between the formulation of the current two-dimensional Stokes flow problem and 
the one utilized in the corresponding injection and suction situations in the classical Saffman-Taylor problem~\cite{Pat,Mir1,Thome}. In the usual Saffman-Taylor instability, the 
flow is governed by Darcy's law, meaning that pressure gradients are balanced by the viscous drag at the walls of the Hele-Shaw cell. However, in the present Stokes flow case, 
the pressure gradients are balanced by viscous forces generated by shear in the flow plane. Furthermore, in the Saffman-Taylor problem, the flow is potential, and only 
admits two boundary conditions, imposed for the normal component of the stress, and velocity fields \cite{Batchelor,Langlois,Mir1, Nagel}. On the other hand, in our current 
problem the flow is not potential and also, we must specify boundary conditions at the interface for both the tangential and normal components of the velocity, and stress 
fields~\cite{Batchelor,Langlois,Tanv1, Nagel}. These differences make the development of a perturbative, second-order mode-coupling theory significantly more laborious in the 
two-dimensional Stokes flow case.

During the injection or suction process, the initially unperturbed, circular interface can become unstable, and deform, due to the interplay of viscous 
and capillary forces. Within the scope of our perturbative scheme, we write the perturbed interface as $ \mathcal{R}(\theta,t) = R(t) + \zeta(\theta,t)$, where 
\begin{equation}\label{radius}
R(t) = \sqrt{ R_{0}^{2} + \frac{Q}{\pi}t}
\end{equation}
is the time-dependent radius of the unperturbed interface, with $R_{0}=R(t=0)$ representing the radius of the initial circular boundary. In addition, the net interface perturbation 
is Fourier expanded as
\begin{equation}\label{zetafourier}
\zeta(\theta, t) = \sum_{n=-\infty}^{\infty} \zeta_n(t) \exp{\left(i n \theta\right)},
\end{equation}
where $\zeta_{n}(t)$ stands for the complex Fourier perturbation amplitudes, with integer wave numbers $n$. Note that our perturbative weakly nonlinear approach requires 
that $|\zeta(\theta,t)| \ll R(t)$. To ensure global mass conservation, the zeroth term of Eq.~\eqref{zetafourier} satisfies
\begin{equation}
\zeta_{0}(t) = - \frac{1}{R}\sum_{n \neq 0} |\zeta_{n}(t)|^2.
\end{equation}
Keep in mind that in this work we focus on the weakly nonlinear dynamic regime, and include terms up to second-order in $\zeta$. Linear stability analysis includes only 
linear terms in $\zeta$. Therefore, within such a linear approximation the Fourier modes decouple, so in the end the perturbation is restricted to a single mode. In contrast, 
our second-order weakly nonlinear approach considers the coupling of a full spectrum of Fourier modes. Although we are only at second-order, mode interaction makes a big difference, 
allowing one to get useful information about the pattern's morphologies already at the lowest nonlinear level. In fact, the inclusion of the second-order perturbative terms is essential 
to properly capture and describe the underlying fingering process. Therefore, in this section, our main goal is to derive a system of mode-coupling differential equations that describe the 
time evolution of the interfacial amplitudes $\zeta_{n}(t)$.

In our two-dimensional Stokes flow problem, the base state corresponds to an expanding or retracting circular interface of radius given by Eq.~\eqref{radius}, such that 
the associated velocity fields are just a source/sink
\begin{equation}
   \mathbf{u}^{(0)}_j = \frac{Q}{2\pi r}\mathbf{\hat{r}}.
\end{equation}
From Eqs.~\eqref{stokes1}-\eqref{stokes2} we can obtain that the base state pressures in the fluids are 
uniform, and given by
\begin{equation}
   p^{(0)}_1 = \sigma \frac{1}{R}, \qquad p^{(0)}_2 = 0.
\end{equation}

To obtain the velocities and pressures related to the perturbed interface evolution, as we did in Eq.~\eqref{zetafourier} for the interface perturbation amplitudes, we need to 
write Fourier expansions for the pressure and velocity fields. In this way, we expand the pressure fields as
\begin{equation}\label{pex}
  p_{j} = p^{(0)}_{j} + \sum_{n \neq 0} \tilde{p}_{n,j}(r,t) \exp{\left(i n \theta\right)},
\end{equation}
and the radial velocity fields as
\begin{equation}\label{urad}
  u_{r,j} = \frac{Q}{2\pi r}+ \sum_{n \neq 0}  \tilde{a}_{n,j}(r,t)\exp{\left(i n \theta\right)}.
\end{equation}
Likewise, the azimuthal velocity fields are written as
\begin{equation}\label{uthet}
  u_{\theta,j} = \sum_{n \neq 0}  \tilde{b}_{n,j}(r,t)\exp{\left(i n \theta\right)}.
\end{equation}

In the search for the mode-coupling equations for $\zeta_{n}(t)$, we proceed by taking the divergence of Eq.~\eqref{stokes1}, and using Eq.~\eqref{stokes2}, to find that 
the pressure is harmonic
\begin{equation}\label{lapla}
  \nabla^2p_j= 0.
\end{equation}
By plugging in the pressure expansion \eqref{pex} into Laplace's equation~\eqref{lapla}, and Fourier transforming, yields
\begin{equation}\label{pcoe}
  \tilde{p}_{n,j} =  p_{n,j} \left(\frac{r}{R}\right)^{(-1)^{j + 1}|n|}.
\end{equation}

Then, to obtain the radial velocity Fourier amplitudes appearing in Eq.~\eqref{urad}, we substitute Eq.~\eqref{pex} and Eq.~\eqref{pcoe} back into the radial 
component of Eq.~\eqref{stokes1}, and take the Fourier transform, arriving at
\begin{equation}
\frac{d^2 \tilde{a}_{n,j}}{dr^2} +  \frac{3}{r}\frac{d \tilde{a}_{n,j}}{dr} + \frac{(1 - n^2)}{r^2} \tilde{a}_{n,j} = (-1)^{j + 1}\frac{|n|}{R}\frac{p_{j,n}}{\eta_j} \left(\frac{r}{R}\right)^{(-1)^{j + 1}|n|-1}.
\end{equation}
The solution of the above equation is 
\begin{equation}\label{a}
 \tilde{a}_{n,j} = \frac{|n|}{|n| + (-1)^{j + 1}}\frac{p_{n,j} R}{4\eta_j} \left(\frac{r}{R}\right)^{(-1)^{j + 1}|n| + 1} + u_{n,j}\left(\frac{r}{R}\right)^{(-1)^{j + 1}|n| - 1}.
\end{equation}
Note that Eq.\eqref{a} is constituted by an inhomogenous part, proportional to the pressure fields, and by a homogeneous part which depends on the coefficients $u_{n,j}$.

To have access to the azimuthal components given by Eq.~\eqref{uthet}, we substitute Eq.~\eqref{urad} and Eq.~\eqref{a} into the incompressibility condition [Eq.~\eqref{stokes2}], 
to get
\begin{equation}\label{b}
 \tilde{b}_{n,j} = i\frac{(-1)^{j + 1}|n| + 2}{n}\frac{|n|}{|n| + (-1)^{j + 1}}\frac{p_{n,j} R}{4\eta_j} \left(\frac{r}{R}\right)^{(-1)^{j + 1}|n| + 1} + i\frac{(-1)^{j + 1}|n|}{n} u_{n,j}\left(\frac{r}{R}\right)^{(-1)^{j + 1}|n| - 1}.
\end{equation}

For each Fourier mode $n$, we must obtain four coefficients $p_{n,1}$, $p_{n,2}$, $u_{n,1}$ and $u_{n,2}$. To do that, we expand the boundary conditions given by Eqs.~\eqref{bc12}-\eqref{bc4}, the normal and tangent unit vectors in Eq.~\eqref{rt}, retaining terms up to order $\zeta^2$
\begin{equation}
    \boldsymbol{\hat{n}} \approx \left(1 - \frac{1}{2}\frac{\zeta^2_{\theta}}{R^2} \right)\mathbf{\hat{r}} - \frac{\zeta_{\theta}}{R}\left(1 - \frac{\zeta}{R} \right)\boldsymbol{\hat{\theta}} \qquad     \boldsymbol{\hat{t}} \approx  \frac{\zeta_{\theta}}{R}\left(1 - \frac{\zeta}{R} \right)\mathbf{\hat{r}} + \left(1 - \frac{1}{2}\frac{\zeta^2_{\theta}}{R^2} \right)\boldsymbol{\hat{\theta}},
\end{equation}
and do the same for the curvature term appearing in \eqref{bc4},
\begin{equation}
\label{curva}
\kappa = \nabla\cdot\boldsymbol{\hat{n}} \approx \frac{1}{R} -\frac{1}{R^2}\left( \zeta + \zeta_{\theta\theta}\right) + \frac{1}{R^3}\left( \zeta^2 + \frac{1}{2}\zeta^2_{\theta}  + 2\zeta\zeta_{\theta\theta}\right).
\end{equation}
Plugging Eqs.\eqref{pex}, \eqref{urad}, \eqref{uthet}, \eqref{pcoe}, \eqref{a}, and \eqref{b} into the boundary conditions, consistently expanding the equations up to the second-order in $\zeta$ using the above relations, and Fourier transforming, after considerable calculation and manipulation, we arrive at an algebraic problem for the unknown coefficients. 
The considerably long expressions for the coefficients  $p_{n,1}$, $p_{n,2}$, $u_{n,1}$, and $u_{n,2}$ are presented in the appendix~\ref{app}. By substituting these resulting 
expressions for the coefficients into the kinematic boundary condition [Eq.~\eqref{kin}], we finally obtain the equation of motion for the perturbation amplitudes (for $n \ne 0$)
\begin{equation}\label{result}
    \Dot{\zeta}_n = \lambda(n) \zeta_n + \sum_{n' \neq 0} \mathcal{T}(n,n') \zeta_{n'}\zeta_{n-n'},
\end{equation}
where the overdot represents a total time derivative, and
\begin{equation}\label{growth}
    \lambda(n)  = -\left[\frac{Q}{2\pi R^2} + \frac{\sigma}{2(\eta_1 + \eta_2) R}|n|\right]
\end{equation}
denotes the linear growth rate. This expression for $\lambda(n)$ agrees with the linear dispersion relation calculated in Refs.~\cite{Amar1,Amar2,Sayag2}. In addition, 
the second-order mode-coupling term is given by
\begin{eqnarray}\label{tee}
    \mathcal{T}(n,n')  = -\frac{1}{2}\bigg[\frac{Q}{\pi R^3}\left(A|n|{\rm sgn}[n'(n - n')] - 1\right) - \frac{\sigma}{2(\eta_1 + \eta_2) R^2}\bigg(|n'| + |n - n'| - |n|{\rm sgn}[n'(n - n')]\bigg)\bigg],
\end{eqnarray}
with the sign function ${\rm sgn}$ being equal to $\pm 1$ according to the sign of its argument. Equation~\eqref{result}, which is a central result of this work, is the mode-coupling equation of the two-dimensional Stokes problem for injection and suction.

\section{Discussion}
\label{discuss}

Prior to addressing the nonlinear aspects of the two-dimensional Stokes problem under injection and suction which are related to the morphology of the pattern-forming 
structures, in Sec.~\ref{lin} we analyze certain fundamental features of the linear theory. This analysis is informative, and also useful for our subsequent weakly nonlinear 
investigation which will be performed in Sec.~\ref{non}.

\subsection{Linear regime}
\label{lin}

We begin by discussing the linear growth rate expression given by Eq.~\eqref{growth}. By inspecting the second term on the right hand side of Eq.~\eqref{growth}, one can 
readily see that, as usual, surface tension tends to stabilize interface disturbances. On the other hand, the first term on the right hand side of 
Eq.~\eqref{growth} is somewhat peculiar. Notice that this term assumes positive values, only if we have suction, i.e. if $Q<0$. Consequently, the 
interface is linearly unstable for suction, but stable for injection, when $Q>0$. Somewhat surprisingly, this suction-injection linear 
behavior for two-dimensional Stokes flows is independent of the viscosity contrast $A$. It should be emphasized that this last feature is in stark contrast to what 
happens in the traditional viscous fingering (VF) problem in radial Hele-Shaw cells, where the linear dispersion relation is written as~\cite{Pat,Mir1}
\begin{equation}
\label{lambdapat}
\lambda_{\rm VF}(n)=\frac{Q}{2 \pi R^{2}} (A |n| - 1) - \frac{\sigma b^{2}}{12(\eta_{1} + \eta_{2})R^{3}} |n| (n^{2} - 1),
\end{equation}
where $b$ denotes the thickness of the Hele-Shaw cell. Note that the modes $n = 0$ and $n = \pm 1$ correspond to a dilation of a circular interface, and to 
a global off-center shift of the circular interface, respectively. So, if one is concerned with a perturbed (i.e., noncircular) interface, the acceptable modes would 
be $|n| \ge 2$. From the first term on the right hand side of Eq.~\eqref{lambdapat} it is evident that the viscosity contrast $A$ has a key role in determining the linear stability 
of the interface. In the Darcy's law regulated, radial viscous fingering problem, interfacial instability can occur for injection (suction) if $A>0$ ($A<0$). Therefore, 
irrespective of the sign of $Q$, the interface can only deform if the displacing fluid has smaller viscosity. 

The linear growth rate for the two-dimensional Stokes flow [Eq.~\eqref{growth}] is also unusual in another aspect: it decreases linearly with mode $n$, in such a way that 
the mode of maximum growth rate is
\begin{equation}
\label{maxstokes}
n_{\rm max}=\pm 2,
\end{equation}
corresponding to the smallest mode leading to interface deformation. Thus, assuming all modes to be initially present, and with comparable amplitudes, the resulting interface shape should exhibit only two lobes. However, an experimental realization of a two-dimensional Stokes flow with suction in a biological system does not corroborate such an exotic linear prediction~\cite{Amar2}. As a matter of 
fact, the establishment of a selection mechanism which determines the ultimate number of fingers formed in a two-dimensional Stokes flows is still an open, and challenging 
problem. There are some suggestions for such mode selection in the literature, but they are not consensual~\cite{Amar1,Amar2}. Nevertheless,  in previous theoretical studies 
of the two-dimensional Stokes flow problem, the symmetry of the interface was imposed through the initial conditions~\cite{Tanv1,Tanv2}. We note that experimentally ``preparing" 
the initial conditions to have a certain symmetry is possible~\cite{Lacalle}.

The linear prediction for two-dimensional Stokes flow expressed by Eq.~\eqref{maxstokes} is also in contrast with the equivalent result for the radial viscous fingering case, 
in which the mode of maximum growth rate is obtained by setting $ [d \lambda_{\rm VF}(n) / dn ]_{n=n_{\rm max}^{\rm VF}} = 0$, yielding
\begin{equation}
\label{maxvf}
n_{\rm max}^{\rm VF}=\pm \sqrt{\frac{1}{3} \left [ 1 + \frac{6QAR (\eta_{1} + \eta_{2} )}{\pi \sigma b^{2}} \right ]}
\end{equation}
which is a time-dependent quantity since $R=R(t)$. 

Another important linear quantity that can be extracted from Eq.~\eqref{growth} is the critical (or, marginal) mode [obtained by setting $\lambda(n)=0$]
\begin{equation}
\label{critstokes}
n_{\rm crit}=-\frac{Q (\eta_{1} + \eta_{2} )}{\pi \sigma R}.
\end{equation}
Considering the unstable situation $Q<0$, this is the mode beyond which all other modes are linearly stable. Note that, as $R$ becomes smaller, the number of unstable mode increases. This mechanism is also at odds with the usual Saffman-Taylor instability, for which this cascade of modes happens as $R$ increases \cite{Mir1}.

From the material discussed above, it is clear that the mechanism leading to interface destabilization in two-dimensional Stokes flow is substantially 
different from the one responsible for triggering the Saffman-Taylor instability. Therefore, we now present a brief discussion on the mechanism leading 
to interface instabilities in the two-dimensional Stokes flow situation. It turns out that the instability mechanism can be trivially understood. For suction, 
the base flow (unperturbed) velocity is directed toward the origin, and its magnitude increases as the interface radius decreases ($v\sim 1/r$). When the 
interface is perturbed away from a circle, points located closer to the origin are drawn radially inward more strongly than points farther from the origin. 
As a consequence, the interface tends to become more deformed. For injection, the opposite process takes place, and the interface tends to become more stable 
as it evolves outward. 

A similar type of process is also present in the Saffman-Taylor instability problem. However, it only constitutes a secondary 
mechanism. The dominant mechanism for the onset of the Saffman-Taylor instability is the different viscous resistance of the two fluids at the walls of the 
Hele-Shaw cell. Thus, since the resistance depends on the fluids' viscosities, the usual viscous fingering instability is primarily dependent on the viscosity 
contrast. As mentioned in Sec.~\ref{intro}, wall resistance is negligible in the two-dimensional Stokes flow situation considered in this work. Thus, 
the viscosity contrast does not interfere in the linear stability of the interface. Nonetheless, as we will see in Sec.~\ref{non} the viscosity contrast $A$ still 
substantially  affects the interface morphology. In order to tackle these important morphological effects, one needs to go beyond the linear regime, and explore the interface 
dynamics at the onset of the nonlinear stage of the evolution. Surprisingly, our lowest-order weakly nonlinear approach is capable to capture the most relevant 
aspects of the interface morphology.

\subsection{Nonlinear stage}
\label{non}
 
We initiate our discussion about the weakly nonlinear regime of the two-dimensional Stokes by calling the readers' attention to a very important point. If on one hand, 
it is true that the linear growth rate $\lambda(n)$ [Eq.~\eqref{growth}] does not depend on the viscosity contrast $A$, on the other hand, it also a fact that the second-order 
mode-coupling function $\mathcal{T}(n,n')$ [Eq.~\eqref{tee}] does depend on $A$. The verification that such an important dependence on $A$ arises already at the lowest nonlinear 
level is promising. It opens up the possibility of investigating the role played by $A$ in determining the shape of the interfacial patterns already at second-order. Therefore, 
our major goal in this section is to use the mode-coupling equations~\eqref{result}-\eqref{tee} to obtain perturbative solutions for the two-dimensional Stokes flow interfacial patterns 
in the case of suction ($Q<0$). 

By considering the coupling of a finite number of participating Fourier modes, we aim to extract the most important morphological features of the emerging patterns, and possibly 
get perturbative pattern-forming structures resembling near-cusp fingering shapes obtained in Refs.~\cite{Tanv1,Tanv2,Poz,How3,Cum2}. Recall that 
such almost-cusp shapes have been accessed via exact solutions, or numerical simulations based on conformal mapping techniques. It should be stressed that, contrary to studies 
based on conformal mappings, whose results are restricted to cases in which $A=-1$ or $A=1$, our mode-coupling theory can address a whole range of allowed values for 
the viscosity contrast, i.e. $-1 \le A \le 1$. Another advantage of our scheme is the fact that it is perturbative on the interface deformation $\zeta$, but 
nonperturbative on the surface tension $\sigma$.

With the mode-coupling equation~\eqref{result} at hand, it is relatively simple to obtain the time evolution of fluid-fluid interfaces. To generate the two-dimensional Stokes 
flow patterns, we consider the nonlinear coupling of a finite number $N$ of Fourier modes. More specifically, we consider the coupling of a fundamental mode $n$ and its harmonics 
$2n$, $3n$, ${\ldots}$, $Nn$, and rewrite Eq.~\eqref{result} in terms of the real-valued cosine amplitudes $a_{n}(t) = \zeta_{n}(t) + \zeta_{-n}(t)$. Then, the growing patterns 
are generated by numerically solving the corresponding coupled nonlinear differential equations for the mode amplitudes $a_{n}(t)$. Once this is done, the shape of the interface 
is found by utilizing Eq.~\eqref{zetafourier}. In addition, to make sure that the interfacial behaviors we detect are spontaneously induced by the weakly nonlinear dynamics, 
and not by artificially imposing large initial amplitudes for the harmonic modes, we always set the initial ($t = 0$) harmonic mode amplitudes to zero. Therefore, at $t=0$ only 
the fundamental mode $n$ has a nonzero, but small amplitude. This is done to avoid artificial growth of the harmonic modes, imposed solely by the initial conditions.

\begin{figure*}
	\includegraphics[width=6.0 in]{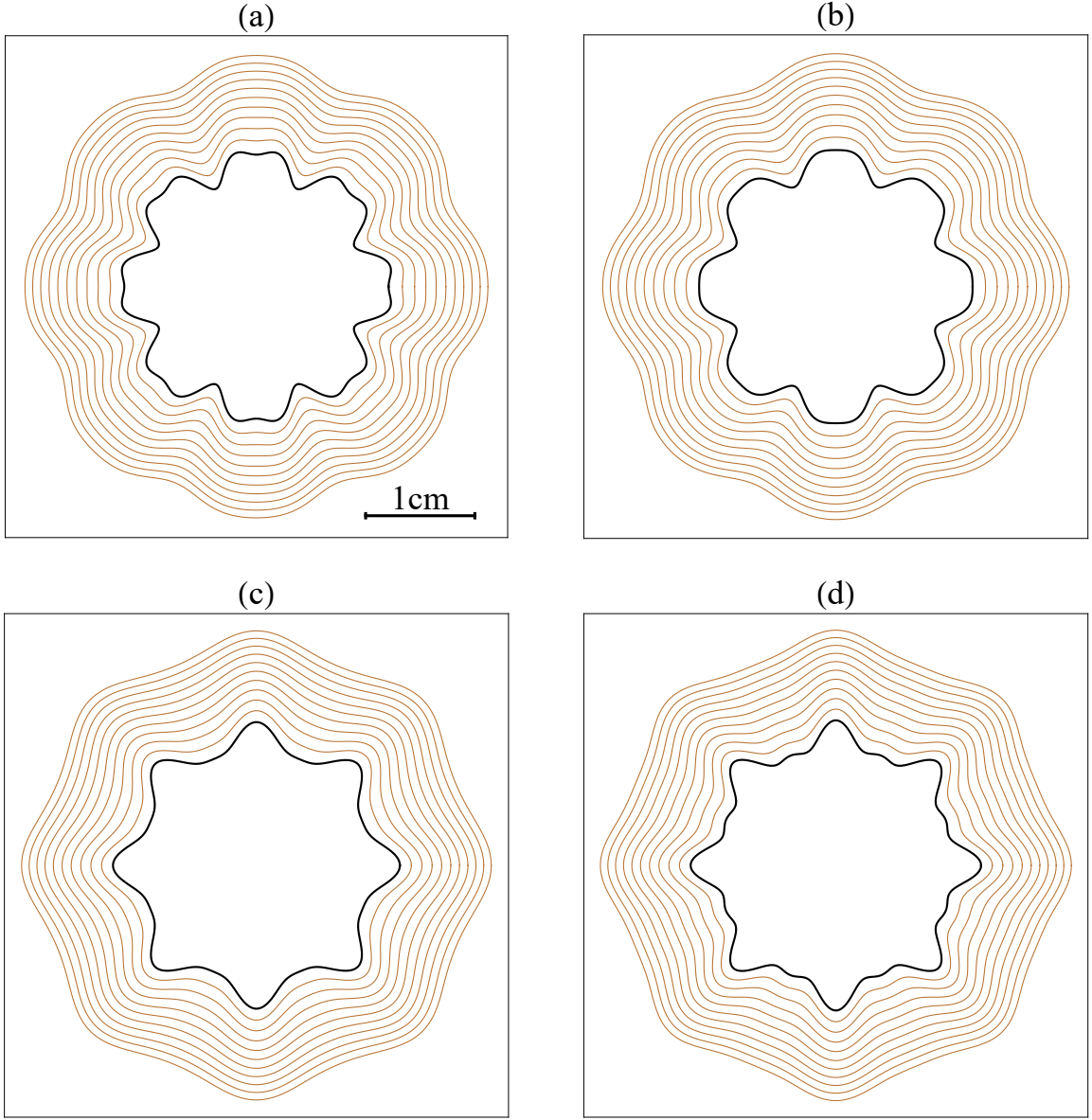}
	\caption{Representative weakly nonlinear time evolution of the fluid-fluid interface for two-dimensional Stokes flow with suction. The patterns are obtained by considering 
		the coupling of two Fourier modes $n$, and $2n$. The values of the viscosity contrast are (a) $A=-1$, (b) $A=-0.5$, (c) $A=0.5$, and (d) $A=1$. The innermost 
		interfacial pattern (taken at final time $t=t_{f}=49.8$ s) is highlighted, being represented by a ticker curve. The horizontal bar in (a) indicates 1 cm.}
	\label{patterns2}
\end{figure*}

\begin{figure*}
	\includegraphics[width=6.0 in]{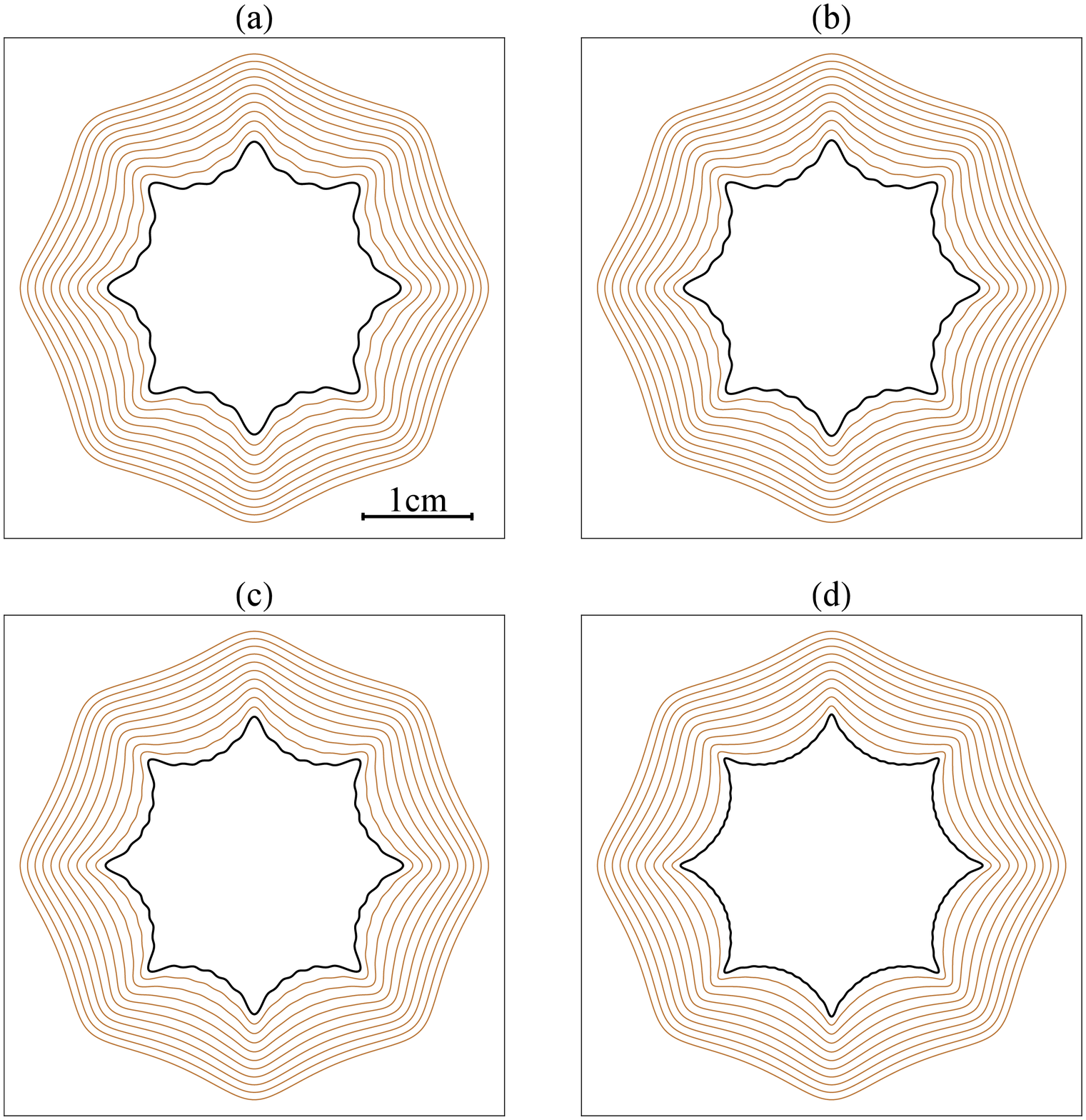}
	\caption{Weakly nonlinear evolution of the two-dimensional Stokes flow patterns for viscosity contrast $A=1$. These patterns are generated by using a increasingly larger 
		number of participating modes $N$: (a) 3, (b) 4, (c) 5, and (d) 10. The innermost interfacial pattern (taken at final time $t=t_{f}=49.8$ s) is highlighted, 
		being represented by a ticker curve.The horizontal bar in (a) indicates 1 cm.}
	\label{many}
\end{figure*}

\begin{figure*}
	\includegraphics[width=5.5 in]{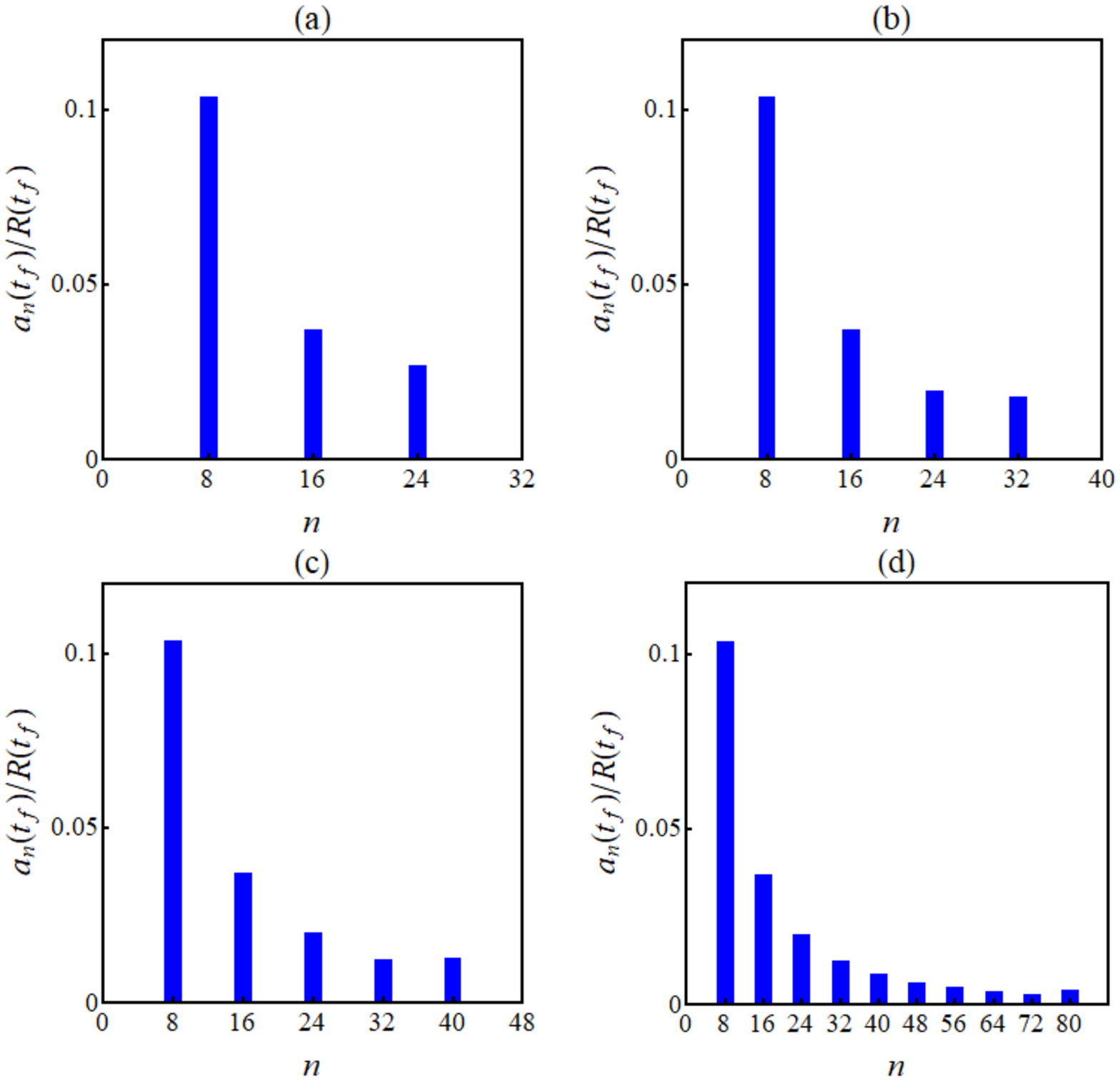}
	\caption{Rescaled cosine mode amplitudes $a_{n}(t)$/$R(t)$ taken at the final time $t=t_{f}$, as a function of the azimuthal mode number $n$, when 
		(a) $N=3$, (b) $N=4$, (c) $N=5$, and (d) $N=10$. The data represented by the vertical bars are extracted from the corresponding patterns that appear in Fig.~\ref{many}.
	Note that for the largest amplitude (mode $n=8$) we have that $a_{8}(t_{f})$ $\approx 10,5 \%$ of $R(t_{f})$.}
	\label{histo}
\end{figure*}

\begin{figure*}
	\includegraphics[width=6.0 in]{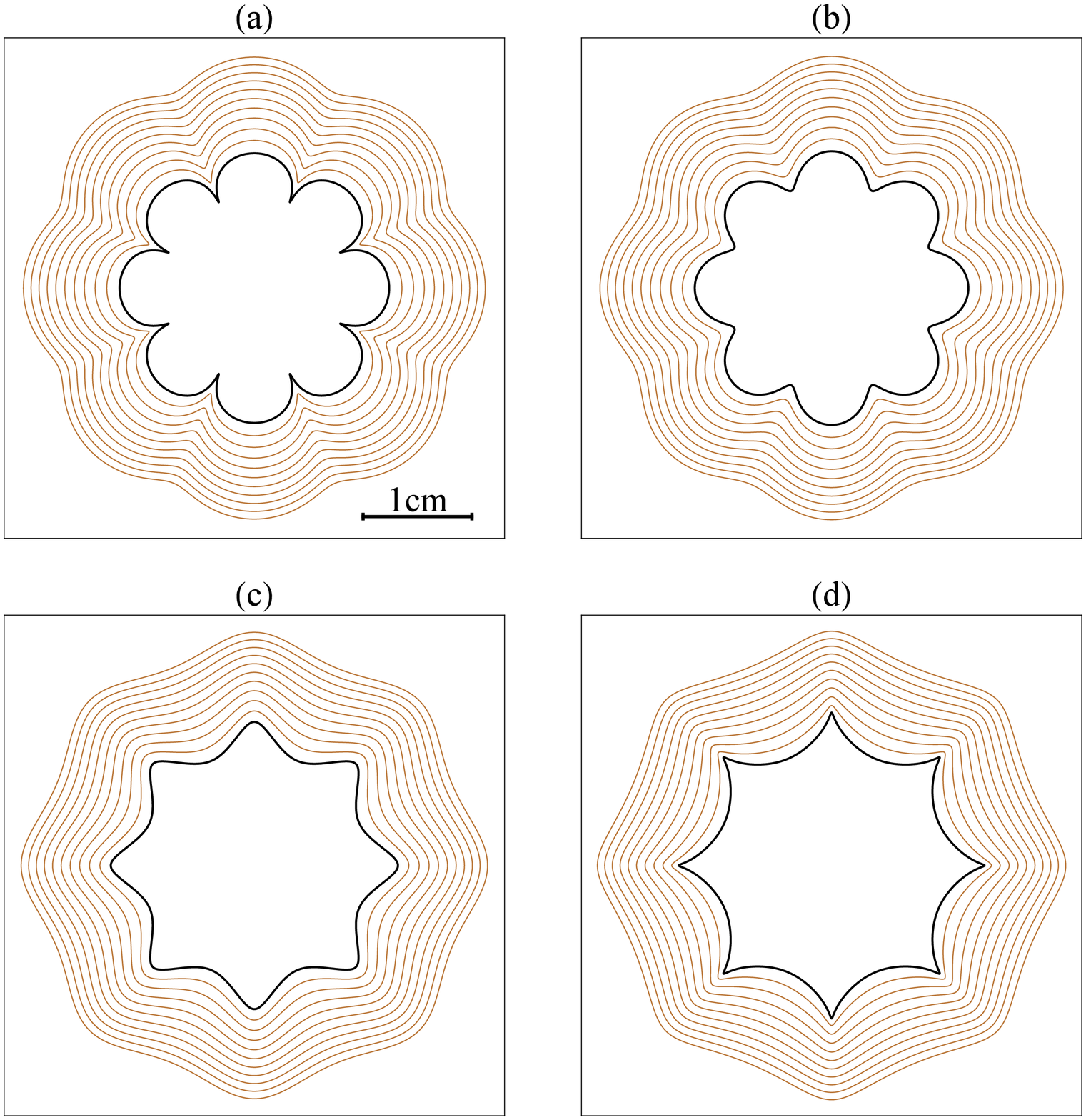}
	\caption{Representative weakly nonlinear time evolution of the fluid-fluid interface for two-dimensional Stokes flow with suction. The patterns are obtained by considering 
		the coupling of sixty participating modes ($N=60$). The values of the viscosity contrast are (a) $A=-1$, (b) $A=-0.5$, (c) $A=0.5$, and (d) $A=1$. As customary, the innermost 
		interfacial pattern (taken at final time $t=t_{f}=49.8$ s) is highlighted, being represented by a ticker curve. The horizontal bar in (a) indicates 1 cm.}
	\label{patterns60}
\end{figure*}

\begin{figure*}
	\includegraphics[width=5.5 in]{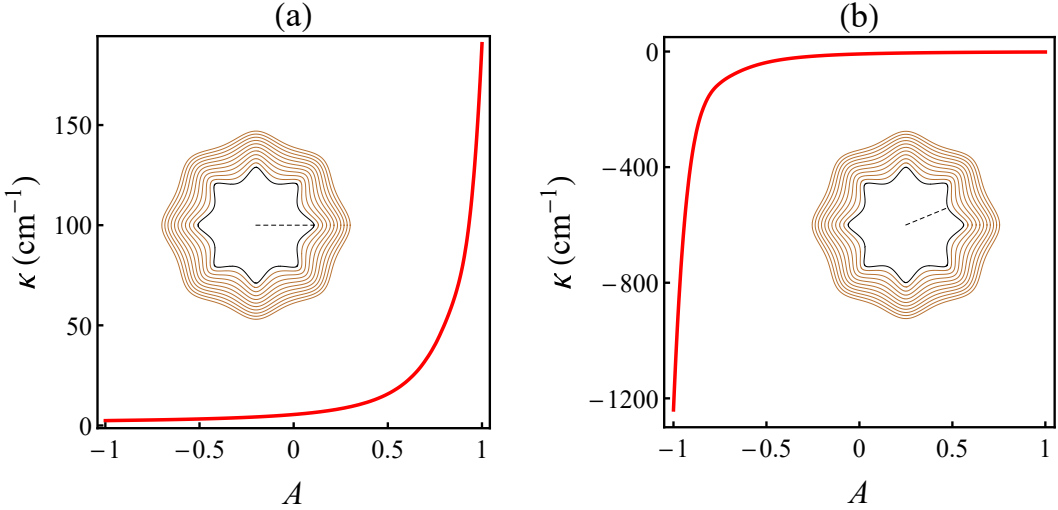}
	\caption{Behavior of the interface curvatures $\kappa$ as the viscosity contrast $A$ is varied from $-1$ to $1$, 
		for two specific angular locations: (a) $\theta=0$, and (b) $\theta= \pi / 8$. To guide the eye, the insets locate these angles for the 
		pattern depicted in Fig.~\ref{patterns60}(c) for $A=0.5$. Here $\kappa$ is computed at the final time $t=t_{f}=49.8$ s as in Fig.~\ref{patterns60}.}
	\label{curvatures}
\end{figure*}

\begin{figure*}
	\includegraphics[width=6.5 in]{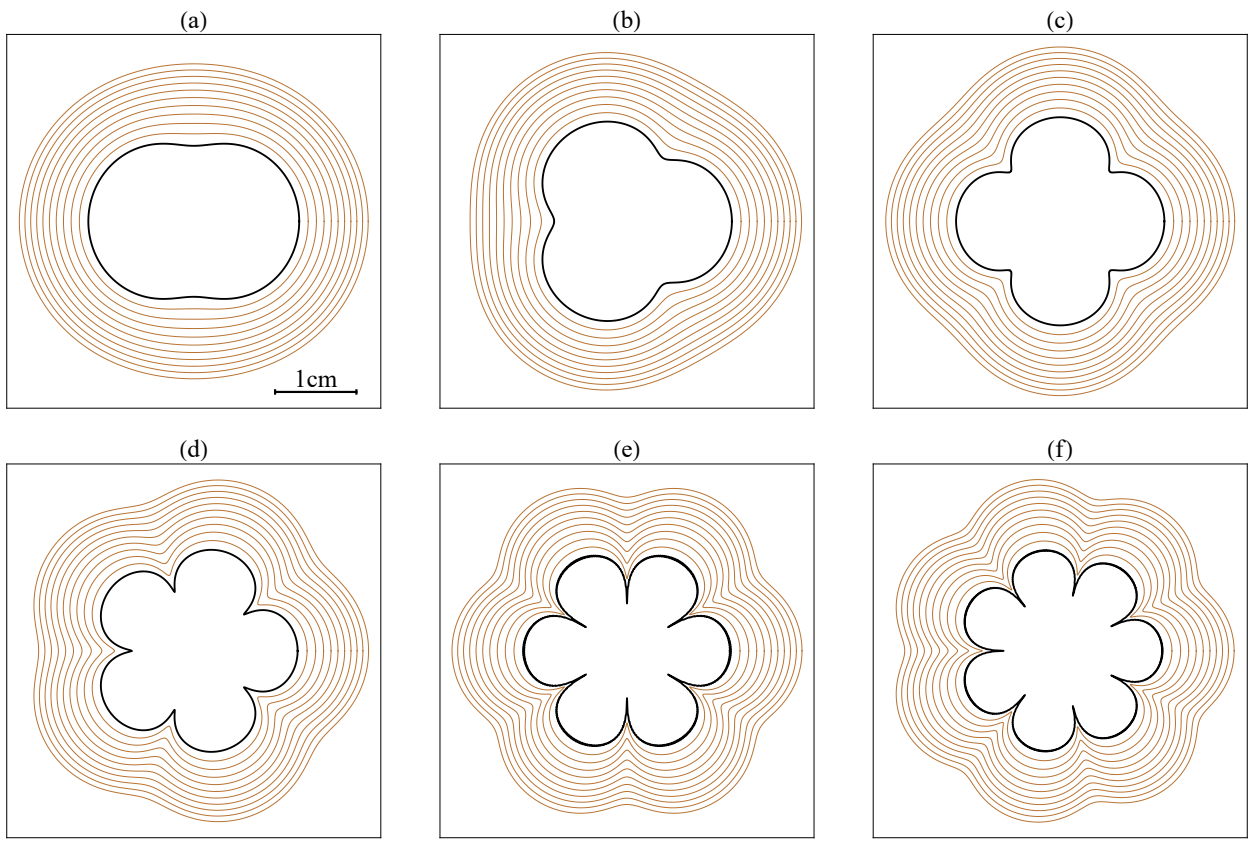}
	\caption{Representative weakly nonlinear time evolution of the interface for two-dimensional Stokes flow with suction, with 
		viscosity contrast $A=-1$. The patterns are obtained by considering that the fundamental mode, and its initial amplitude are 
		taken as (a) $n=2$, and $a_{n}(0)= R_{0}/450$ cm; (b) $n=3$, and $a_{n}(0)= R_{0}/350$ cm; (c) $n=4$, and $a_{n}(0)= R_{0}/280$ cm; 
		(d) $n=5$, and $a_{n}(0)= R_{0}/210$ cm; (e) $n=6$, and $a_{n}(0)= R_{0}/160$ cm; (f) $n=7$, and $a_{n}(0)= R_{0}/150$ cm. All the other 
		parameters are identical to those used in Figs.~\ref{patterns2}-\ref{curvatures}. In (a)-(f) the innermost interfacial pattern is 
		highlighted, being represented by a ticker curve. The horizontal bar in (a) indicates 1 cm. In (a)-(e) for the largest mode amplitude 
		we have that $a_{n}(t_{f})$ $\approx 15 \%$ of $R(t_{f})$, while in (f) $a_{n}(t_{f})$ $\approx 12.5 \%$ of $R(t_{f})$.}
	\label{patterns30b}
\end{figure*}

In order to strengthen the practical and academic relevance of our weakly nonlinear results, throughout this work we use typical parameter values for all physical quantities 
that are in line with the ones utilized in existing experimental and theoretical studies in radial Hele-Shaw cells~\cite{Pat,Chen,Cardoso,Mir1,Praud,Li,Bischo,Thome}. In 
this way, in the various situations analyzed in this work we consider that the viscosities of the fluids may vary within the range $0 \le \eta_{j} \le 10$ {\rm g/cm~s}. The 
suction rate is taken as $Q=-9~{\rm cm^{2}/s}$, and the surface tension between the fluids is $\sigma=0.6~{\rm dyne/cm}$. In addition, we evolve from the initial radius 
$R_{0}=12~{\rm cm}$, and consider the initial amplitude of the fundamental mode as $a_{n}(0)=R_{0}/150~{\rm cm}$. In Figs.~\ref{patterns2}-\ref{curvatures} we take, without loss of generality, the fundamental mode as $n=8$. The effect of considering a different value for the fundamental $n$ on the resulting patterns will be 
discussed at the end of this section, in the analysis of Fig.~\ref{patterns30b}. Finally, we point out that in all calculations and plots presented in this work, we paid close attention to the limit of validity of our perturbative theory, in such a way that we always make sure that $|\zeta_{n}(t)| \ll R(t)$.

We initiate our analysis by examining the simplest second-order scenario, and consider the interplay of only two Fourier modes ($N=2$): the fundamental $n=8$, and its first 
harmonic $2n=16$. The choice of using precisely these modes (the fundamental $n$, and its first harmonic $2n$) to begin our investigation can be justified as follows. 
Remember that an emblematic feature of the bubble ($A=1$) and blob ($A=-1$) shape solutions obtained via complex variable techniques in Refs.~\cite{Tanv1,Tanv2,Poz,How3,Cum2} 
is the formation of very sharp, cusplike fingers. It turns out that a two-mode mode-coupling approach is quite appropriate to examine pattern-forming mechanisms involving the growth of 
sharp fingering structures. It has been shown that finger-tip sharpening, widening, and splitting are behaviors related to the influence of a fundamental mode $n$ on the
growth of its harmonic mode $2n$~\cite{Mir1}. In fact, within the scope of mode-coupling, these fundamental pattern formation phenomena can be predicted, captured 
and properly described already at second-order in the perturbation amplitudes. 

The occurrence of these types of finger tip phenomena in two-dimensional Stokes flow with suction can be examined by rewriting the mode-coupling equation~\eqref{result} 
for cosine mode amplitudes, and analyzing the influence of a fundamental mode $n$ on the growth of its harmonic $2n$. In this situation, the equation of motion for the growth 
of the harmonic mode amplitude can be written as
\begin{equation}
\label{twomodes}
 \Dot{a}_{2n} = \lambda(2n) a_{2n} + \frac{1}{2} \mathcal{T}(2n,n)  a_{n}^{2},
\end{equation}
where
\begin{equation}
\label{teetwomodes}
\mathcal{T}(2n,n)= - \bigg[\frac{Q}{\pi R^3}\left( A|n| - \frac{1}{2} \right) \bigg].
\end{equation}
Note that the mode-coupling function $\mathcal{T}(2n,n)$ assumes a pretty simple form that depends on the viscosity contrast $A$. The interesting 
point about the function $\mathcal{T}(2n,n)$  is that it controls the finger shape behavior, and ultimately the morphology of 
the resulting pattern. The sign of $\mathcal{T}(2n,n)$ dictates whether finger tip-sharpening or finger tip-widening is favored by the dynamics. 
From Eq.~\eqref{twomodes} we see that if $\mathcal{T}(2n,n) >0$, the result is a driving term of order  $a_{n}^{2}$ forcing growth of 
$a_{2n} >0$, the sign that is required to cause inward-pointing fingers to become wide, favoring finger tip-broadening. In contrast, if
$\mathcal{T}(2n,n) <0$ growth of $a_{2n} <0$ would be favored, leading to inward-pointing finger tip-sharpening.

In order to further investigate the analytic predictions for the finger tip shape behavior provided by Eqs.~\eqref{twomodes} and~\eqref{teetwomodes}, in Fig.~\ref{patterns2} 
we plot representative two-dimensional flow patterns generated by considering the coupling of modes $n=8$ and $2n=16$. This is done for four values of the viscosity contrast: 
(a) $A=-1$ ($\eta_{1}=10$~{\rm g/cm~s} and $\eta_{2}=0$); (b) $A=-0.5$ ($\eta_{1}=7.5$~{\rm g/cm~s} and $\eta_{2}=2.5$~{\rm g/cm~s});(c) $A=0.5$ ($\eta_{1}=2.5$~{\rm g/cm~s} and 
$\eta_{2}=7.5$~{\rm g/cm~s}); and (d) $A=1$ ($\eta_{1}=0$ and $\eta_{2}=10$~{\rm g/cm~s}). The interfaces are plotted for times $48.8~{\rm s} \le t \le 49.8~{\rm s}$ in intervals 
of 0.1 s. Note that while varying the values of the viscosity contrast $A$, we keep the sum $\eta_{1} + \eta_{2}$ fixed in order to maintain the linear growth rate~\eqref{result} 
unchanged as $A$ is modified. These very same values of $A$, and times will be used to plot Figs.~\ref{many} and~\ref{patterns60}. By inspecting Fig.~\ref{patterns2} it is clear 
that the inward moving interface becomes increasingly unstable as time progresses. One interesting point is that, despite the fact that the linear growth rate $\lambda(n)$ does 
not depend on $A$, the morphology of the final patterns (represented by the innermost interfaces) do depend on the value $A$. Note that if $A<0$ (or, if $\mathcal{T}(2n,n) <0$)
[Figs.~\ref{patterns2}(a)-\ref{patterns2}(b)] one observes the formation of inward moving fingers that are sharp, while the structures separating these sharp fingers look blunt 
[Fig.~\ref{patterns2}(b)], or split into two small protuberances [Fig.~\ref{patterns2}(a)]. The opposite behavior is observed if $A>0$ (or, if $\mathcal{T}(2n,n) >0$) 
[Figs.~\ref{patterns2}(c)-\ref{patterns2}(d)] where the inward moving finger look wider (or even split at their tips), while the structures separating consecutive wide finger look 
sharp. Of course, these pictorial observations are in agreement with the analytical predictions provided by our discussion of Eqs.~\eqref{twomodes} and~\eqref{teetwomodes}.

By examining Fig.~\ref{patterns2} we also note that as the viscosity contrast varies from $A=-1$ to $A=1$, the tips of the inward moving fingers located at polar angle 
$\theta=\pi /8$ tend to become wider, while the interface point located at angle $\theta=0$ tends to get locally sharper. In this sense, our two mode second-order mimic of the 
pattern formation dynamics indicates that sharp fingers pointing inward should arise at $\theta=\pi /8$ in the blob case ($A=-1$), while sharp structures pointing 
outward should appear at $\theta=0$ in the bubble case ($A=1$). These last observations are reassuring in the sense that they are in qualitative agreement with what has been 
observed in Refs.~\cite{Tanv1,Tanv2,Poz,How3,Cum2} regarding the formation of near-cusps. However, the weakly nonlinear patterns depicted in Fig.~\ref{patterns2} are 
still very different from the cusped interfaces generated in Refs.~\cite{Tanv1,Tanv2,Poz,How3,Cum2}.

In the pursuit of getting a closer morphological similarity between our weakly nonlinear patterns, and the nice cusplike shapes obtained in Refs.~\cite{Tanv1,Tanv2,Poz,How3,Cum2}, 
we pass to consider the coupling of an increasingly larger number $N$ of participating Fourier modes. This is done in Fig.~\ref{many}, where we investigate the influence
of the number of participating Fourier modes in determining the shape of the two-dimensional Stokes flow patterns at second-order. In Fig.~\ref{many} we focus on the case 
in which $A=1$, and produce patterns using the coupling of modes $n$, $2n$, ${\ldots}$, $Nn$, where $n=8$ is the fundamental mode: (a) $N=3$, (b) $N=4$, (c) $N=5$, and (d) $N=10$. It should be noted that all other physical quantities and initial conditions are exactly the same as those used in Fig.~\ref{patterns2}.

By going through Fig.~\ref{many}, it is apparent that the consideration of a larger number of interacting Fourier modes leads to promising results. The weakly nonlinear patterns 
portrayed Fig.~\ref{many} reveal the formation of eightfold polygonal-like morphologies (set by the fundamental mode) having concave-shaped edges. These edges present small undulations determined by the number of participating modes $N$. Since the fluid-fluid interface is not discontinuous, no Gibbs phenomenon is present, and the Fourier series converges. Therefore, as $N$ is increased, the amplitude of the oscillations decreases, and the edges of the polygonal patterns look increasingly smoother. This is clearly illustrated by 
Figs.~\ref{many}(a)-\ref{many}(d). Eventually, for a sufficiently large $N$ these oscillations are no longer observed. It is also noticeable that the vertices of the patterns 
become sharper as $N$ is augmented. Finally, it is worthwhile to note that the weakly nonlinear pattern depicted in Fig.~\ref{many}(d) for $N=10$ bears a close resemblance to the
typical near-cusp shapes obtained by Tanveer and Vasconcelos via complex variable methods in Refs.~\cite{Tanv1,Tanv2}.

Complementary information about the pattern-forming structures illustrated in Fig.~\ref{many} is provided by Fig.~\ref{histo} which plots the value of the cosine Fourier
amplitudes $a_{n}(t)$, rescaled by the unperturbed radius of the interface $R(t)$ (given by the vertical bars), at the final time $t=t_{f}=49.8$ s, for various 
participating modes $n$. From the Fourier spectra shown in Fig.~\ref{histo} it is evident that the cosine mode amplitudes drop very quickly as $n$ is 
increased. Despite this quick drop, the Fourier series does not converge as rapidly, and the consideration of a greater number of modes is necessary to make the edges of the polygonal patterns to become sufficiently smooth.

Inspired by the findings of Fig.~\ref{many} and Fig.~\ref{histo}, in Fig.~\ref{patterns60} we display a representative set weakly nonlinear two-dimensional Stokes flow patterns 
with suction, now using a sufficiently large number of participating modes $(N=60)$ in such a way that the edges of the polygonal-like patterns are quite smooth. Similar to 
what we did in Fig.~\ref{patterns2}, the patterns displayed in Fig.~\ref{patterns60} are obtained for four values of the viscosity contrast: (a) $A=-1$, (b) $A=-0.5$, 
(c) $A=0.5$, and (d) $A=1$. All the rest of the physical parameters and initial conditions are equal to the ones used in Fig.~\ref{patterns2}. Figure~\ref{patterns60} 
is quite elucidating since it allows one to visualize the impact of the viscosity contrast $A$ on the shape of the emerging patterns, already at the lowest nonlinear level. 
For example, the weakly nonlinear pattern shown in Fig.~\ref{patterns60}(a) for $A=-1$ reveals the most salient features encountered in the complex-variable-generated 
shapes obtained for the sucking of a blob of fluid (see, for instance, Fig. 3 in Ref.~\cite{Cum2}). In Fig.~\ref{patterns60}(a) we have an eightfold polygonal-like 
morphology, now having convex-shaped edges that meet at near-cusp, inward-pointing indentations (see, for example, the near-cusp formed at $\theta = \pi / 8$). On the other hand, when 
$A=-0.5$ [Fig.~\ref{patterns60}(b)] no near-cusp indentations are found, and the inward-pointing fingers are not that sharp. 

In addition, when $A=0.5$ [Fig.~\ref{patterns60}(c)] yet another type of pattern arises: it has a starfishlike shape, and also does not show any signs of near-cusp formation. 
By contrasting the finger shapes in illustrated in Figs.~\ref{patterns60}(b)-\ref{patterns60}(c), one sees that while the fingering structures formed at $\theta = \pi / 8$ 
become more rounded and wider, the ones produced at $\theta = 0$ tend to become sharper and narrower. This trend persists when we look at the pattern portrayed in 
Fig.~\ref{patterns60}(d) for $A=1$: in $\theta = \pi / 8$ the fingering structure has a near-circular shape, whereas in $\theta = 0$ a near-cusp, outward-pointing 
finger is unveiled. As anticipated by the structure obtained in Fig.~\ref{many}(d) for $N=10$, the pattern displayed in Fig.~\ref{patterns60}(d) for $N=60$ is indeed an 
eightfold polygonal-like interface, but now it has smooth, concave-shaped edges. Incidentally, the weakly nonlinear structure represented in Fig.~\ref{patterns60}(d) 
does have all essential morphological elements of the typical shapes obtained for the sucking of a bubble in Refs.~\cite{Tanv1,Tanv2} (see, for instance, Fig. 4 
in~\cite{Tanv1}, and Fig. 1 in~\cite{Tanv2}).

To better substantiate the impact of the viscosity contrast $A$ on the morphologies of the two-dimensional Stokes flow patterns for a whole range of values 
of $A$, in Fig.~\ref{curvatures} we plot the value of the interface curvatures $\kappa$ at time $t=t_{f}=49.8$ s, as $A$ is varied from $A=-1$ to $A=1$, for points located at two 
important angular locations along the interface: (a) at $\theta=0$, and (b) $\theta=\pi /8$. The interface curvature can be readily calculated from Eq.~\eqref{curva}.
Note that the rest of the parameters and initial conditions used in Fig.~\ref{curvatures} are equal to those utilized in Fig.~\ref{patterns2}.
By observing Fig.~\ref{curvatures}(a) we see that the interface curvature at $\theta=0$ is relatively small for negative values of $A$, and then starts to grow very 
significantly as $A$ varies from 0 to 1, reaching a maximum value at $A=1$, where a near-cusp fingered structure is formed. Note that in (a) $\kappa >0$ 
meaning the fingers at $\theta=0$ point outward. On the other hand, in Fig.~\ref{curvatures}(b) we verify that at $\theta=\pi /8$, the curvature is very 
high and negative for $A=-1$ where an inward-pointing near-cusp finger arises, and then becomes much less intense as the viscosity contrast changes from $A=-1$ 
to $A=0$, until assuming considerably small negative values as $A \rightarrow 1$. Observe these quantitative remarks about the behavior of $\kappa$ with $A$ 
in Fig.~\ref{curvatures} are in accordance with the more visual verifications one can make at angles $\theta=0$ and $\theta=\pi /8$ in the patterns depicted 
in Fig.~\ref{patterns60}. The dramatic variation of $\kappa$ with $A$ illustrated in Fig.~\ref{curvatures} reinforces the importance of the viscosity contrast 
in determining the overall morphology of the pattern-forming structures in our problem.

We close this section by discussing the effect of considering a different value for the fundamental mode $n$ on the generated two-dimensional 
Stokes flow, weakly nonlinear patterns with suction. In Figs.~\ref{patterns2}-\ref{curvatures}, we took $n=8$. As mentioned earlier in this work, the choice 
of $n=8$ as the fundamental mode was made without loss of generality. It turns out that $n=8$ is a linearly unstable mode for the time interval 
used in Figs.~\ref{patterns2}-\ref{curvatures}. If one chooses another linearly unstable Fourier mode as being the fundamental mode, the basic 
physical results are similar to the ones obtained for $n=8$. This is illustrated in Fig.~\ref{patterns30b} which shows suction patterns 
produced for viscosity contrast $A=-1$, by taking different values for the fundamental mode and its initial amplitude, namely for 
(a) $n=2$, and $a_{n}(0)= R_{0}/450$ cm; (b) $n=3$, and $a_{n}(0)= R_{0}/350$ cm; (c) $n=4$, and $a_{n}(0)= R_{0}/280$ cm; (d) $n=5$, 
and $a_{n}(0)= R_{0}/210$ cm; (e) $n=6$, and $a_{n}(0)= R_{0}/160$ cm; (f) $n=7$, and $a_{n}(0)= R_{0}/150$ cm. Other than that, these patterns 
are generated by utilizing all the physical parameters used in Figs.~\ref{patterns2}-\ref{curvatures}. By inspecting Fig.~\ref{patterns30b} 
one readily observes that the types of pattern-forming structures obtained when $A=-1$ for these modes of lower wave number than $n=8$ are morphologically 
similar to the corresponding structure shown in Fig.~\ref{patterns60}(a) when $n=8$ and $A=-1$. In other words, all these structures for $A=-1$ and 
different $n$'s have a characteristic polygonal-like morphology, having convex-shaped edges that can meet at near-cusps, inward-pointing dents. We have 
verified that this also occurs for all other values of $A$.

\section{Concluding remarks}
\label{conclude}

Most of the studies of the fingering pattern formation in two-dimensional Stokes flow with suction focus either on the linear stability 
analysis of the early time dynamics, or on the use of complex variable techniques for the exploration of advanced time stages of 
the flow. These investigations concentrate their attention basically on two extreme situations: (i) extraction of a bubble of negligible 
viscosity, for which the viscosity contrast $A=1$; and (ii) contraction of a blob of viscous fluid, for which $A=-1$. In this work, we examined 
different aspects of the problem: through the employment of a perturbative, second-order mode-coupling theory, we aim attention at the 
weakly nonlinear intermediate stages of the flow that bridge purely linear and fully nonlinear regimes. Additionally, instead of examining 
just the cases for $A=-1$ or $A=1$, we explored a whole range of permitted values of $A$ (i.e., $-1 \le A \le 1$). Our analytical approach 
provides useful insights into the basic mechanisms involved in the pattern formation process, and into the role of the viscosity contrast in 
determining the shape of the emerging fingering structures.

As expressed by the equation of motion for the problem [Eq.~\eqref{result}], at the linear level, second-order mode-coupling reproduces linear 
stability results. Accordingly, we derived an expression for the linear growth rate of the two-dimensional Stokes flow problem. This linear dispersion 
relation reveals some peculiarities if compared with its equivalent expression for the classical Saffman-Taylor problem in radial Hele-Shaw cells. The 
most salient difference is that, as opposed to what happens in the Saffman-Taylor case, the growth rate for two-dimensional Stokes flow is independent 
of the viscosity contrast $A$. Therefore, linearly, $A$ plays no role in the stability of the interface. Different aspects of the linear growth rate have been discussed, 
a physical mechanism for explaining the Stokes flow instability has been provided, and some other useful information about the linear stability of the system 
have been extracted. It should be pointed out that our linear growth rate [Eq.~\eqref{growth}] coincides with the expression previously derived in 
Refs.~\cite{Amar1,Amar2,Sayag2}. This agreement supports the validity and correctness of our mode-coupling calculation.

At the weakly nonlinear level, by utilizing Eq.~\eqref{result}, we have shown that our second-order perturbative interfacial solutions do a decent job in reproducing the typical 
near-cusp morphologies conventionally obtained by analytical and numerical studies based on complex variable methods when $A=1$ and $A=-1$. Moreover, we 
have found that for $-1 < A < 1$, a whole series of unexplored interfacial patterns arise, but at second-order these structures do not display 
the occurrence of near-cups. Our nonlinear findings also demonstrate that $A$ has a key role in determining the finger-tip curvature behavior of both inward and 
outward pointing fingers. In summary, our perturbative results show that, despite its unimportant value for the 
linear dynamics, the viscosity contrast is vital in setting the dynamics and shape of interfacial patterns in two-dimensional Stokes flow with suction. 
It is fortunate, and a bit surprising, that all these relevant aspects about the morphology of the two-dimensional Stokes 
flow patterns can be caught and predicted already at the lowest nonlinear level. 

\begin{acknowledgments}
J. A. M. thanks CNPq (Conselho Nacional de Desenvolvimento Cient\'{\i}fico e Tecnol\'ogico) for financial support under Grant No. 305140/2019-1. G. D. C. also 
thanks CNPq for financial support of a postdoctoral fellowship under contract No. 150959/2019-2. We gratefully acknowledge important discussions with Carles Blanch-Mercader and Giovani Vasconcelos. We are indebted to Carles Blanch-Mercader for a critical reading of the manuscript, and also for providing detailed comments
and valuable suggestions. 
\end{acknowledgments}

\appendix*
\section{Expressions for $p_{n,1} $, $p_{n,2} $, $u_{n,1}$, $u_{n,2}$}
\label{app}
This appendix presents the expressions for the pressure and velocity Fourier amplitudes which appear in the text.

\begin{eqnarray}
p_{n,1} &=&   \eta_1(n + {\rm sgn}(n)) \frac{(n -{\rm sgn}(n)) \pi R \sigma + 2 {\rm sgn}(n) Q (\eta_1 - \eta_2)}{\pi R^3 (\eta_1 + \eta_2)}\zeta_n \nonumber \\ &-& \eta_1\sum_{n'\neq0}\bigg[(1 + |n|)\bigg( ((n - n')^2 + n'^2 +|n| - |(n - n')n'| -2) \frac{\sigma}{R^3 (\eta_1 + \eta_2)}   \nonumber\\ &+& 2\eta_2( 3 + 3|n|-2|n-n'| -2|n'|)\frac{(\eta_1 - \eta_2) Q}{\pi R^4 (\eta_1 + \eta_2)^2}\nonumber \\ &+& 2\eta_1( 3 - |n|+2|n-n'| +2|n'|)\frac{(\eta_1 - \eta_2) Q}{\pi R^4 (\eta_1 + \eta_2)^2}\bigg)\bigg]\zeta_{n-n'}\zeta_{n'},
\end{eqnarray}
\begin{eqnarray}
p_{n,2} &=&   -\eta_2(n - {\rm sgn}(n)) \frac{(n +{\rm sgn}(n)) \pi R \sigma - 2 {\rm sgn}(n) Q (\eta_1 - \eta_2)}{\pi R^3 (\eta_1 + \eta_2)}\zeta_{n}\nonumber \\ &-& \eta_2\sum_{n'\neq0}\bigg[(-1 + |n|)\bigg( ((n - n')^2 + n'^2 -|n| - |(n - n') n'| -2) \frac{\sigma}{R^3 (\eta_1 + \eta_2)}   \nonumber\\ &+& 2\eta_1( 3 - 3|n|+2|n-n'|+2|n'|)\frac{(\eta_1 - \eta_2) Q}{\pi R^4 (\eta_1 + \eta_2)^2}\nonumber \\ &+& 2\eta_2( 3 +|n|-2|n-n'| -2|n'|)\frac{(\eta_1 - \eta_2) Q}{\pi R^4 (\eta_1 + \eta_2)^2}\bigg)\bigg]\zeta_{n-n'}\zeta_{n'},
\end{eqnarray}
\begin{eqnarray}
u_{n,1} &=&  -\frac{n(({\rm sgn}(n) + n) \pi R \sigma + 2 {\rm sgn}(n) Q (\eta_{1} - \eta_{2}))}{4\pi R^2 (\eta_1 + \eta_2)}\zeta_{n} \nonumber \\ &+& \sum_{n'\neq0}\bigg[n \bigg(((n - n')^2 + n'^2 +|n| - |(n - n') n'|)\frac{\sigma}{4 R^2 (\eta_1 + \eta_2)} \nonumber\\ &+& 2\eta_1(  1 -|n|+2|n-n'| +2|n'|)\frac{(\eta_1 - \eta_2) Q}{\pi R^3 (\eta_1 + \eta_2)^2}\nonumber \\ &+& 2\eta_2( 1 + 3|n|-2|n-n'| -2|n'|)\frac{(\eta_1 - \eta_2) Q}{\pi R^3 (\eta_1 + \eta_2)^2}\bigg)\bigg]\zeta_{n-n'}\zeta_{n'},
\end{eqnarray}
\begin{eqnarray}
u_{n,2} &=&  \frac{n((-{\rm sgn}(n) + n) \pi R \sigma - 2 {\rm sgn}(n) Q (\eta_{1} - \eta_{2}))}{4\pi R^2 (\eta_1 + \eta_2)}\zeta_{n} \nonumber \\ &+& \sum_{n'\neq0}\bigg[n \bigg(((n - n')^2 + n'^2 -|n| - |(n - n') n'|)\frac{\sigma}{4 R^2 (\eta_1 + \eta_2)} \nonumber\\ &+& 2\eta_2(  1 +|n|-2|n-n'| -2|n'|)\frac{(\eta_1 - \eta_2) Q}{\pi R^3 (\eta_1 + \eta_2)^2}\nonumber \\ &+& 2\eta_1( 1 - 3|n|+2|n-n'| +2|n'|)\frac{(\eta_1 - \eta_2) Q}{\pi R^3 (\eta_1 + \eta_2)^2}\bigg)\bigg]\zeta_{n-n'}\zeta_{n'}.
\end{eqnarray}

\end{document}